\documentclass[conference]{IEEEtran}

\usepackage{amsmath}
\usepackage{color}
\usepackage{multirow}
\usepackage{multicol}
\usepackage{url}
\usepackage{lastpage,algorithm,algorithmic}
\usepackage{epsfig,amssymb,url}
\usepackage{bbding}

\newcommand{\bs}{\boldsymbol}

\newcommand{\ds}{\displaystyle}

\begin{document}
\title{When Cellular Meets WiFi in  Wireless Small Cell Networks}
\author{\IEEEauthorblockN{Mehdi Bennis$^\dag$, Meryem Simsek$^\ddag$,  Walid Saad$^*$,  Stefan Valentin$^\star$, Merouane Debbah$^+$, and Andreas Czylwik$^\ddag$}
\IEEEauthorblockA{$^\ddag$Chair of Communication Systems, University of Duisburg-Essen, Germany\\
$^\star$Bell Labs, Alcatel-Lucent, Stuttgart, Germany \\
$^\dag$Centre for Wireless Communications, University of Oulu, Finland\\
$^*$Electrical and Computer Engineering Department, University of Miami, Coral Gables, FL, USA \\
$^+$SUPELEC, Gif-sur-Yvettes, France\\
Email: \{simsek, czylwik\}@nts.uni-due.de, stefan.valentin@alcatel-lucent.com, bennis@ee.oulu.fi, merouane.debbah@supelec.fr}
}

\maketitle
\vspace{-100ex}

\begin{abstract}
The deployment of small cell base stations~(SCBSs) overlaid on existing macro-cellular systems is seen as a key solution  for offloading traffic, optimizing coverage, and boosting the capacity of future cellular wireless systems.
The next-generation of SCBSs is envisioned to be \emph{multi-mode}, i.e., capable of transmitting   \emph{simultaneously}  on both licensed and unlicensed bands. This constitutes  a cost-effective integration of both WiFi and cellular radio access technologies (RATs) that can efficiently cope with  \emph{peak} wireless data traffic and heterogeneous quality-of-service requirements. To leverage the advantage of such multi-mode SCBSs, we discuss the novel proposed paradigm of  \emph{cross-system learning}  by means of which  SCBSs self-organize  and autonomously steer their traffic flows across different RATs. Cross-system learning allows the SCBSs to leverage the advantage of both the WiFi and cellular worlds. For example, the SCBSs can offload delay-tolerant data traffic to WiFi, while simultaneously learning  the  probability distribution function of their  transmission strategy over the licensed cellular band. This article will first introduce the basic building blocks of cross-system learning and then provide preliminary performance evaluation in a Long-Term Evolution (LTE) simulator overlaid with WiFi hotspots. Remarkably, it is shown that the proposed cross-system learning approach  significantly outperforms a number of benchmark traffic steering policies.
 \end{abstract}
\IEEEpeerreviewmaketitle


\section{Introduction}
Owing to the proliferation of sophisticated mobile devices (i.e., smartphones, tablets),  a $20$-fold increase in  data traffic is expected over the next few years, compelling mobile operators  to find new ways to significantly boost their network capacity, provide better coverage, and reduce network congestion  \cite{Juniper}. In this context, the idea of  heterogeneous networks (HetNets), consisting of a mix of short-range and low-cost small cell base stations (SCBSs) underlaying the macrocell network, has recently emerged as a key solution for solving this  capacity crunch problem \cite{QC}. However, reaping the potential benefits of heterogeneous and small cell  deployments is contingent upon: a) developing innovative interference management, load balancing, and traffic offloading mechanisms, and b) integrating different radio access technologies (RATs), tiers (femto-, pico-, micro-, metro-, and macro- cells), and licensed/unlicensed frequency bands.  
\begin{figure}[!t]
	\centering
		\includegraphics[width=0.45\textwidth]{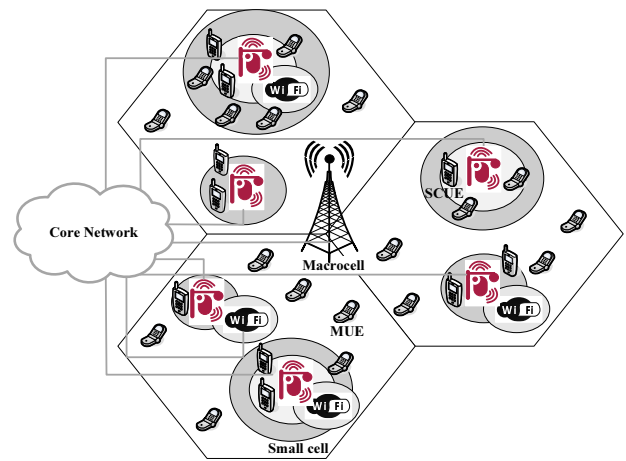}
	\caption{An illustration of a macrocell deployment underlaid with multi-mode small cell base stations. Gray areas refer to the cell range expansion bias per SCBS.}
	\label{fig:HetNetWifi}
\end{figure}

An efficient, cost-effective integration of cellular (e.g., 3G/LTE) and WiFi technologies, referred to as \emph{inter-RAT}, has recently attracted significant interest from academia, industry, and standardization bodies alike \cite{Juniper,Lee,Zhuo}. It is envisioned that WiFi and SCBS deployment will exhibit complementary benefits  that can be leveraged for an efficient integration. On the one hand, due to the uncontrolled, unlicensed nature of WiFi, the competition for resources among a large number of hotspot  users, notably  when other devices (laptops, tablets and dongles) transmit on the same unlicensed band, can yield dramatically poor throughputs. In such scenario, offloading some of this traffic to the well-managed small cell network, operating over the licensed spectrum, can improve the performance. On the other hand, the inherent constraints of small cell networks, particularly due to cross-tier and co-tier interference, motivate offloading some of the traffic to the WiFi band, so as to alleviate the interference and ease  congestion. With the deployment of multi-mode SCBSs operating on both the WiFi and licensed bands, smart traffic offloading strategies that harness the benefits of both worlds must be developed \cite{QC}.

To date, the majority of traffic offloading studies focused on intra-RAT offloading, in which the macrocell traffic is offloaded to smaller cells (e.g.,  femto-, picocells) using cell range expansion (CRE)  and almost blank sub-frame (ABS) \cite{Lopez}. The concept of CRE has been recently proposed, in which a positive range expansion bias  is added to the picocell's pilot downlink received signal strength \cite{QC}. In contrast, the literature on inter-RAT integration, particularly with multi-mode SCBSs is still in its infancy.  To date, WiFi offloading has been studied in a number of works such as in  [4-6]. In  \cite{Lee},  a quantitative study of the performance of $3$G mobile data offloading through WiFi networks is studied. The authors in \cite{Zhuo} proposed a framework for $3$G traffic offloading based on the idea of motivating mobile users with high delay tolerance to offload their traffic to WiFi networks. In \cite{Zhang}, the authors investigate the capacity offload problem among service providers using standard game theoretic tools leading to the inefficient Nash equilibria \cite{Sutton}. Recently, the work  in \cite{QC}  demonstrated that  complementing  heterogeneous cellular networks with WiFi hotspots can be  an attractive solution for operators. A framework to offload traffic between cellular and WiFi RATs was proposed  in \cite{interdigital}. However, this work  focused  on a single WiFi carrier  coexisting with indoor femtocells. Although existing works established the potential of 3G-WiFi coexistence, there is still a need for considerable research  to address pertinent challenges such as  self-organization and dynamic traffic steering between $3$G/LTE and WiFi.

In this article, we introduce a  fully distributed and dynamic traffic offloading framework, in which SCBSs  seamlessly steer their traffic between cellular and WiFi RATs, depending on the traffic type, users' quality-of-service (QoS) requirements,  network load, and interference levels.  SCBSs are assumed to have a wired backhaul connection to the core network, and the impact of heterogeneous backhauls is out of the scope of this article. This developed framework, hereafter coined \emph{cross-system learning}, endows SCBSs with self-organizing capabilities allowing them to \emph{simultaneously} transmit on both $3$G and WiFi bands.  Here, SCBSs carry out a \emph{long-term} optimization procedure by \emph{learning} their \emph{optimal} transmission strategy over  licensed/unlicensed bands, without exchanging information. In particular, delay-tolerant applications can be offloaded to WiFi, when possible, while delay-stringent applications  can be steered towards $3$G/LTE. In contrast to traditional scheduling algorithms (such as  proportional-fair (PF)) which  overlook  users' heterogeneous demands, incorporating a proactive and traffic-aware scheduler is shown to exhibit significant gains,  outperforming a number of benchmark traffic steering policies.

The rest of this article is organized as follows. In Section II, a discussion of the small cell and WiFi paradigms and their  potential integration is presented. In Section III, a basic small cell system model is presented, followed by the  cross-system learning framework for self-organizing radios. A case study along with some numerical results are presented in Section IV, while conclusions are drawn in Section V.

\section{Small Cells and WiFi: A Best of Both World Approach}

Offloading cellular traffic to WiFi and small cells  is seen by operators as a key solution for handling the continuous growth in mobile data traffic.  Offloading traffic to WiFi  will continue to  play a key role due to its low cost-per-bit and sufficient spectrum to support high throughput (notably at $5$ GHz). On the other hand,  operators are able to manage the spectrum used by small cells, optimize their traffic,  and decide where to place them. Nonetheless, as small cells operate on the same spectrum as the macrocell,  coordination between the macro- and small cells  is crucial to mitigate the impact of interference.

Thus far, both cellular and WiFi radio access technologies have been in constant competition, until recently in which  a tighter integration of both technologies has emerged as a necessary paradigm. Indeed, when deployed along each other, operators can not only perform classical offload (through WiFi), but also a smart fine-grained offload, whereby operators can decide which traffic flows over which RAT, while leveraging user's QoS requirements, latency, and backhaul conditions for service differentiation \cite{QC}, \cite{context}. In addition,  with the advent of multi-mode SCBSs, operators can reduce their site acquisition costs (site leasing, installation and backhaul) by combining several RATs into a single device.  This can further lead to a reduction in their capital expenditures (power, memory, etc). So as to reap the benefits of SCBSs' multi-mode capabilities, operators need to devise dynamic offloading strategies and make intelligent decisions aiming at enriching users's QoS and avoiding user churn. For instance, when a UE discovers the presence of WiFi in its vicinity,   delay-tolerant traffic (e.g., web browsing) should be offloaded to WiFi, whereas traffic with more stringent data requirements (e.g., multimedia) would remain on the cellular $3$G/LTE RAT. Furthermore, while traffic offloading at the access level is important, backhaul offload is  yet another important component of the cellular and WiFi integration. Here, operators need to take into account the backhaul conditions and congestion level in their offloading policy before deciding to which RAT a user is offloaded to ensure a better and seamless user experience. $3$GPP has proposed two solutions, namely Selected IP Traffic Offload (SIPTO) and Local IP Access (LIPA) to deal with latency and congestions of traffic flow problems, either through the mobile core or IP network \cite{SIPTO}.  SIPTO supports  an IP traffic offload directly to the internet and away from the mobile core network, to reduce the network load. Nonetheless,  operators must be careful in selecting which traffic to offload, as mobility support for SIPTO traffic can be limited. Under LIPA, IP traffic management is designed to  optimize the traffic destined to a local IP Network locally instead of through the mobile core network.  In what follows, we present the novel paradigm  of cross-system learning  used by self-organizing SCBSs for traffic offload between $3$G/LTE and WiFi.

\section{Cross-System Learning Framework in Self-Organizing Radios}\label{SecSimulations}
In this section, we first provide definitions and notions of reinforcement learning, that will be useful in the sequel. Subsequently, we present the novel framework of cross-system learning.
\subsection{Basic Model}
Consider a  macrocell base station (MBS) operating over a set $\mathcal{S} = \lbrace 1, \ldots, S\rbrace$ of $S$ frequency bands. The macrocell is underlaid with a set $\mathcal{K} = \lbrace 1, \ldots, K \rbrace$ of $K$ SCBSs. Each SCBS is \emph{dual-mode} and can transmit over the licensed and unlicensed spectrum band. Let the downlink transmit power of SCBS $k$ at time $t$ on subband (SB) $s$ be $p_k^{(s)}(t)$ and $p_{k,\max}$  be the maximum transmit power of SCBS $k$.
Let the $S$-dimensional vector  $\bs{p}_k(t)=\left(p_k^{(1)}(t),...,p_k^{(S)}(t)\right)$ denote the power allocation  vector of SCBS $k$ at time $t$.  Let $L_k \in \mathbb{N}$ be the number of discrete power levels of SCBS $k$ and denote by $q^{(\ell,s,b)}_k $ its $\ell$-th transmit power level over SB $s$ when using cell range expansion bias  $b \in \lbrace 1, ..., B\rbrace$.
Thus, the cardinality of the strategy set   $\mathcal{A}_k =  \left\lbrace q_{k}^{(\ell,s,b)}\right\rbrace$ of SCBS $k$ is $N_k = L_k\times S\times B$.  

Due to the mutual co-channel interference and coupling among SCBSs' strategies, the joint interference management and traffic offloading problem can be modeled as a game $\mathcal{G}= \Big(\mathcal{K},\{\mathcal{A}_k\}_{k \in \mathcal{K}},\{\bar{u}_k\}_{k \in \mathcal{K}}\Big)$. Here, $\mathcal{K}$ represents the set of SCBSs,  and the action set $\mathcal{A}_k$ of each SCBS $k$ is the joint set of subband, power levels and CRE bias.
Finally, $\bar{u}_{k}$ denotes the long-term performance metric optimized by  each SCBS $k$, in which, at each time $t$,  each SCBS $k$ chooses its action from the finite set $\mathcal{A}_k$ following a probability distribution $\bs{\pi}_k(t)=\left(\pi_{k,q_k^{(1,1,1)}}(t),...,\pi_{k,q_k^{(L_k,S,B)}}(t)\right)$, and  $\pi_{k,q_k^{(l,s,b)}}=\text{Pr}\left(\boldsymbol{p}_k(t)=q_k^{(l,s,b)}\right)$ is the probability that SCBS $k$ selects  action $q_k^{(l,s,b)}$ at time $t$.


\subsection{Reinforcement Learning}
Reinforcement learning (RL) is an area of machine learning, in which  a number of decision makers or players, having often conflicting objectives, must  be able to make autonomous decisions given limited information, so as to optimize a certain cumulative objective function or reward  \cite{Sutton}. RL has had an impact on a variety of disciplines inclusive of game theory, control theory, operations research, information theory, and genetic algorithms. A key design criterion in RL is to develop strategies that allow players to strike a balance between exploring the network and exploiting their accumulated knowledge.  Recently, RL has received significant interest in the context of self-organizing HetNets, as it allows operators to automate their network in a plug-and-play manner and reducing maintenance costs.

In the context of cellular and WiFi integration, the goal of every SCBS is to devise an intelligent and online learning mechanism to optimize its licensed  spectrum transmission, while at the same time leverage WiFi by offloading delay-tolerant traffic. The developed procedure,  dubbed  cross-system learning,  is rooted in the fact that every small cell optimizes its \emph{long-term} performance metric, as a function of its traffic load, interference levels, and users' heterogeneous traffic requirements. In addition, unlike  standard RL \cite{Sutton}, the cross-system learning procedure allows players to \emph{implicitly} coordinate their transmissions with no information exchange, as well as to leverage the coupling  between  LTE and WiFi, which as will be shown increases the overall network performance and significantly speeds up the convergence.

The  cross-system learning framework is composed of the following  interrelated components:
\begin{itemize}
\item \textbf{Subband selection, power level allocation and cell range expansion bias}: every SCBS learns over time how to select appropriate subbands with their corresponding transmit power levels in both licensed and unlicensed spectrum, in which delay-tolerant traffic is steered toward the unlicensed spectrum. In addition,  every SCBS learns its optimal CRE bias to offload the macrocell traffic to smaller cells.
 \item \textbf{Proactive scheduling}:  Once the small cell acquires its subband, the scheduling decision is traffic-aware taking into account users' heterogeneous QoS requirements (throughput, delay tolerance and latency).
\end{itemize}
\subsection{Subband, Power Level and Cell Range Expansion Bias Selection}
During cross-system learning, every SCBS  minimizes over time its \emph{regret} of selecting strategies yielding lower payoffs, while experimenting other strategies to improve its long-term utility estimation.  The considered \emph{behavioral} assumption is that small cells  are interested in choosing a probability distribution  over their transmission strategies which minimizes the regret, where the  regret of SCBS $k$ for not having played action $q_k^{(\ell{_k},s{{}},b{{}})}$ from $n = 1$ up to time $t$ is defined as:
 \begin{equation}\label{eq:regret}
 r_{k,q_k^{(\ell,s,b)}}(t) = \frac{1}{t}\ds\sum_{n = 1}^t \hat{u}_k\Big(q_k^{(\ell,s,b)},\bs{p}_{-k}(n)\Big) - \tilde{u}_k(n)
  \end{equation}
where $\tilde{u}_k(n)$ is the instantaneous utility   observation (i.e., feedback) of SCBS $k$ at  time $n$, obtained by constantly changing its strategy. In addition, to calculate its regret, every SCBS $k$  estimates its utility function $\hat{u}_k(.,.)$ when taking a given action based on local information.

The rationale of (\ref{eq:regret}) is as follows: if the regret is strictly positive, then SCBS $k$ would have obtained a higher average utility by playing action $q_k^{(\ell,s,b)}$ during all previous time instants, and thus,  the SCBS
$k$ ``regrets'' not having done so. In contrast, if (\ref{eq:regret}) is negative, then SCBS $k$ does not regret its strategy selection. Therefore,  each SCBS needs to strike a balance between choosing actions that yield lower regrets (more often than those with higher regrets), and   playing any of the other actions with a \emph{non-zero} probability.

The  behavioral rule  of every SCBS can be modeled by the probability distribution $\beta_k(\bs{r}^+_{k}(t))$  subject to the maximum transmit power constraints $p_{k,\max}$, where:
\begin{equation}
\begin{array}{ll}
\beta_k(\bs{r}^+_{k}(t)) \in \\ \ds\arg\min_{\bs{\pi}_k }  \Big[\ds\sum_{\bs{p}_k \in \mathcal{A}_k} \pi_{k,\bs{p}_{k}}r_{k,\bs{p}_k}(t) + \frac{1}{\kappa_k} H(\bs{\pi}_{k})\Big],
\label{explore}
\end{array}
\end{equation}
where  $\bs{r}_k^+(t)=\max\left( 0, \bs{r}_k(t)\right)$ denotes the vector of positive regrets, and $H(.)$ represents the Shannon entropy function of the mixed strategy $\bs{\pi}_k$. The \emph{temperature} parameter $\kappa_k >0$ represents the interest of SCBS $k$ to choose other actions. The unique solution to the right-hand-side of the continuous and strictly convex optimization problem in \eqref{explore} is:
\begin{equation}
\label{beta}
\beta_{k,q_k^{(l,s,b)}}(\bs{r}^+_{k}(t)) = \frac{\exp\Big(\kappa_k r^+_{k,q_k^{(l,s,b)}}(t)\Big)} {\ds\sum_{\bs{p}_k \in \mathcal{A}_{k} } \exp\Big(\kappa_k r^+_{k,\bs{p}_k}(t)\Big)},
\end{equation}
where $\beta_{k,q_k^{(l,s,b)}}(\bs{r}^+_{k}(t)) >0$ holds with strict inequality regardless of the regret vector   $\bs{r}_k(t)$.

\begin{figure}
	\centering
		\includegraphics[width=0.4\textwidth]{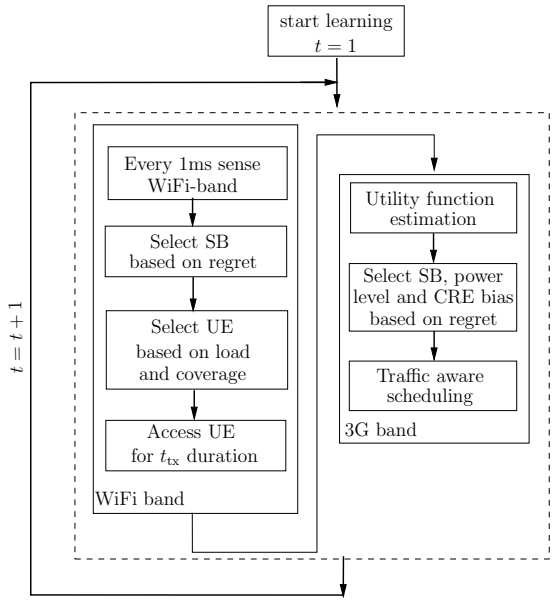}
	\caption{Flow-chart of the proposed cross-system learning procedure performed by each small cell base station.}
	\label{fig:chart}
\end{figure}

Furthermore, given users' different QoS requirements, the cross-system learning framework leverages  WiFi, in which the learning process carried out over WiFi is  \emph{faster} (from a time-scale perspective) than that on the cellular band. More concretely, inspired from the well-known \emph{turbo-principle}, the output (i.e., feedback) from the WiFi learning process is used for the update of the cellular learning process. As will be shown later on, this notion of time-scale significantly reduces the convergence time of the traffic steering algorithm, as compared to standard RL, and  improves the overall performance. Figure \ref{fig:chart} shows a flow-chart of the cross-system learning framework executed by each SCBS.

\subsection{Proactive Scheduling}

Once the SCBSs select their subbands using cross-system learning, they engage in a  proactive and traffic aware scheduling procedure on the selected subband's resource blocks. The  scheduling algorithm is proactive and traffic-aware in nature as it incorporates users' traffic requirements. Notably, the scheduling decision is not only based on the instantaneous channel condition but also on the completion time (delay) and service class of each transmission. For that, within every small cell, all users are sorted in an ascending order as a ratio of their remaining file size and  estimated average data rate.  Then, the  SCBS $k$ computes a metric $D_{k_i}(t)$, which is a function of  the position of  UE $k_i$ and  the number of UEs served by SCBS $k$ at time $t$. Finally,  UE $k_i^*$ is scheduled such that: $ k_i^* =\arg \underset{k_i}{\min} D_{k_i}(t)$.




\section{Cellular and WiFi Offload: A Case Study}
 The developed cross-system learning framework is validated in an integrated LTE-A/ WiFi simulator. The considered scenario  comprises one macrocell consisting of three sectors underlaid with  $K$ open access small cells operating on both $3$G and WiFi. The SCBSs are uniformly distributed within each macro sector, while considering a minimum MBS-SCBS distance of $75$ m. The path-loss models and other set-up parameters were selected according to the $3$rd Generation Partnership Project (3GPP) recommendations for outdoor picocells (model 1) \cite{3GPP}.
$N_{\text{UE}}=30$ mobile UEs were dropped within each macro sector from which $N_{\text{hotspot}} = \frac{2}{3} N_{\text{UE}}/K$ are randomly and uniformly dropped within a $40$ m radius of each SCBS, while the remaining UEs are uniformly dropped within each macro sector. Each UE is assumed to be active, with a fixed traffic model from the beginning of the simulations while moving at a speed of $3$ km/h. The traffic mix consists of different traffic models as shown in Table I, following the requirements of the Next Generation Mobile Networks (NGMN) \cite{NGMN}.
\begin{table}
	\centering	
	\caption{UE traffic mix.}
		\begin{tabular}{|l|l|c|}\hline
		\textbf{Traffic model} & \textbf{Traffic category} & \textbf{Percentage of UEs}\\\hline\hline
		FTP & Best effort & 10\%\\\hline
		HTTP & Interactive & 20\%	\\\hline
		Video streaming & Streaming & 20\%\\\hline
		VoIP & Real-time & 30\%\\\hline
		Gaming & Interactive real-time & 20\%\\\hline
		\end{tabular}
	\label{tab:UETrafficMix}
\end{table}
The bandwidth in the licensed (resp. unlicensed) band is $5$ MHz (resp. $20$ MHz). The simulations are averaged over $500$ transmission time intervals (TTIs). For sake of comparison, we consider the following benchmark algorithms:
\begin{itemize}
\item \emph{\textbf{Macro-only}:}  The macrocell is the only serving cell of all UEs using proportional fair scheduling, by uniformly distributing its maximum transmission power over the whole licensed bandwidth.
\item \emph{\textbf{HetNet}:} the macrocell is augmented with $K$ small cells. Here, both MBS and SCBSs  serve their UEs in the licensed band only, and small cells optimize their subband, power levels and cell range expansion bias.
\item \emph{\textbf{HetNet + WiFi (Load-Based)}:} each SCBS  transmits on both licensed and unlicensed bands by  randomly selecting one subband on both licensed and unlicensed bands. Access to the unlicensed band is performed based on the load as described in Section III.C and Proportional-fair scheduling is performed on the licensed band.
\item \emph{\textbf{HetNet + WiFi (Coverage-Based)}:} Same as \emph{HetNet + WiFi (load-Based)} except that the access method is based on the  reference signal received power (RSRP) criterion.

\end{itemize}
\begin{figure}[h!]
	\centering
		\includegraphics[width=0.47\textwidth]{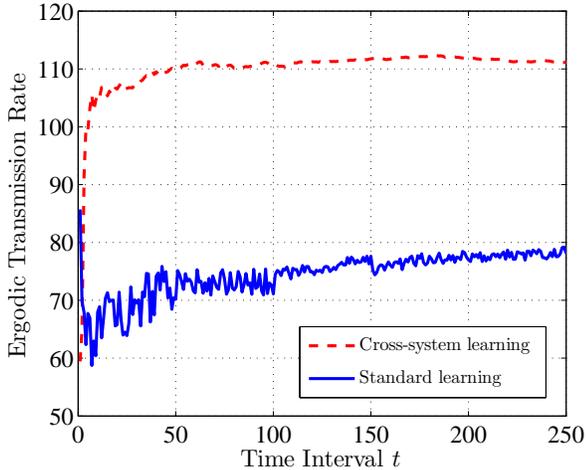}
	\caption{Convergence of the  cross-system learning algorithm vs. the standard independent learning \cite{Sutton}, $K=2$ SCBSs per macrocell sector.}
	\label{fig:ConvergenceRegretICC11}
\end{figure}

\begin{figure}
	\centering
		\includegraphics[width=0.47\textwidth]{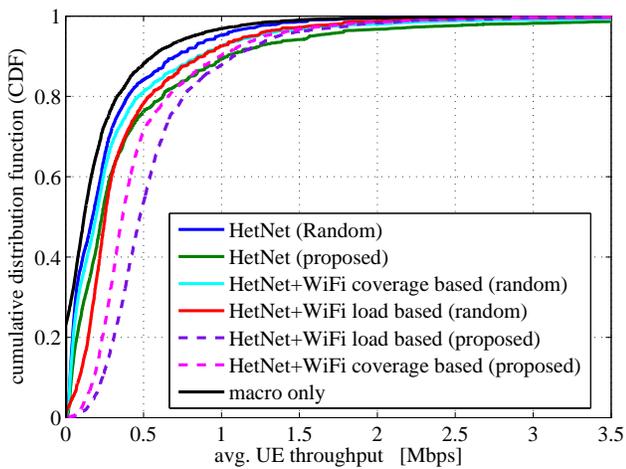}
	\caption{Cumulative distribution function  (CDF) of the average UE throughput for $N_{\text{UE}}=30$ UEs and $K=2$ SCBSs per macrocell sector.}
	\label{fig:UEthroughputBoth}
\end{figure}

\subsection{Convergence}
Figure \ref{fig:ConvergenceRegretICC11} plots the convergence behavior of the  cross-system learning procedure in terms of the ergodic transmission rate (i.e., average cell throughput). Here, we consider $10$ UEs per macro sector, and  $1.4$ MHz bandwidth in the licensed band. In addition, we plot the \emph{standard} RL  algorithm  \cite{Sutton}, in which learning is carried out \emph{independently} over both licensed and unlicensed bands, without any sort of coordination. Quite remarkably, it is shown that the cross-system learning approach converges within less than $50$ iterations, while the standard  approach needs several hundreds iterations to converge. Furthermore, the standard RL procedure exhibits an undesirable oscillating behavior (i.e., ping-pong effects between the licensed and unlicensed band, which can be detrimental in mobility scenarios).

\subsection{Average UE throughput under different offloading strategies}
Figure \ref{fig:UEthroughputBoth} plots the  cumulative distribution function  (CDF) of the average UE throughput for  $N_{\text{UE}}=30$ UEs, and different offloading strategies. Here, random refers to an SCBS which selects randomly one subband and performs PF scheduling,  whereas \emph{proposed} refers to the regret-based learning procedure with traffic-aware  (TA) scheduling.  While, in the \emph{macro-only}  case,  $25\%$ of UEs obtain no rate, deploying small cells on the licensed band  increases the overall performance through suitable cell range expansion bias; especially for cell-edge UEs. The overall performance is further boosted when deploying multi-mode SCBSs  transmitting on both licensed and unlicensed bands (i.e., HetNet+WiFi), particularly for the HetNet+WiFi load-based scenario as compared to the  HetNet+WiFi coverage-based scenario.

\subsection{Impact of scheduling}
\begin{figure}
	\centering
		\includegraphics[width=0.47\textwidth]{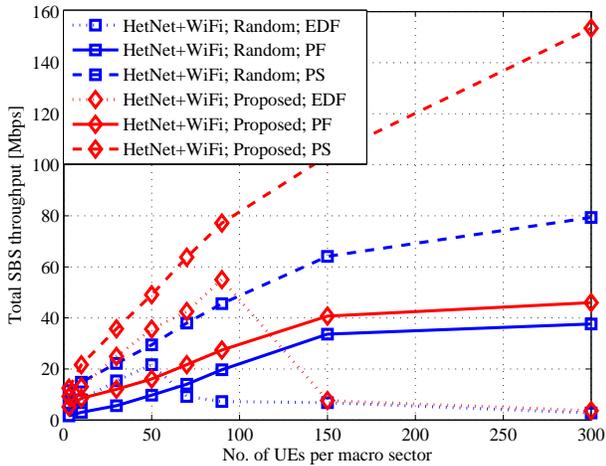}
	\caption{Total cell throughput vs. number of users, for different traffic offloading and scheduling strategies, $K=2$ SCBSs per macrocell sector.}
\label{fig:SBSThrVsIncomingLoad_5}
\end{figure}

Figure \ref{fig:SBSThrVsIncomingLoad_5} shows the total cell throughput as a function of the number of UEs in the network, for the earliest deadline first (EDF), proportional fair (PF), and proactive scheduling (PS) strategies, respectively. While the standard PF scheduler cannot cope with the increasing number of UEs, the traffic-aware scheduling approach judiciously steers users' traffic in an intelligent and dynamic manner over both licensed and unlicensed spectrum, with a $160$-fold increase for  $300$ UEs. These significant gains are rooted in the fact that unlike the proactive scheduler, both EDF and PF schedulers fall short of accounting for the heterogeneous traffic and delay-tolerance nature of their users.

\subsection{Impact of small cell densification}
\begin{figure}[h!]
	\centering
		\includegraphics[width=0.52\textwidth]{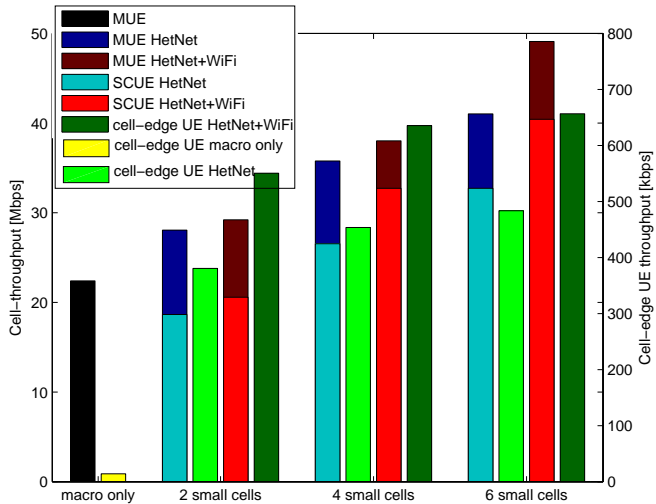}
	\caption{Cell-throughput and cell-edge throughput gains for the ``macro-only'', ``HetNet'', and ``HetNet+WiFi'' offloading strategies, for $K=\lbrace2,4,6\rbrace$ SCBSs.}
	\label{fig:bar}
\end{figure}

Figure \ref{fig:bar} plots the  total cell-throughput  and cell-edge UE throughput for  the macro-only, HetNet, and HetNet+WiFi offloading strategies. Some key observations are worth mentioning. While in the macro-only case, cell edge UEs get rather low throughput gains,  adding $K=2$ small cells is shown to boost users' cell-edge throughput in the HetNet offload case. In addition, a $50\%$ increase in cell-edge UE throughput is obtained with $K=2$ multi-mode small cells (HetNet+WiFi). Furthermore,  small cell users (SCUEs) benefit from the small cells' multi-mode capability when deploying  $K=2$ SCBSs, and this gap further increases when adding more small cells ($K=6$ SCBSs). As a byproduct of this,  offloading is shown to improve not only the performance of SCUEs, but also MUEs, for $K=\lbrace2,4,6\rbrace$ small cell base stations.

%
%

\section{Conclusion}
In this article, we  studied the strategic \emph{coexistence} between $3$G/LTE and WiFi networks in a heterogeneous network, in which multi-mode SCBSs transmit \emph{simultaneously} on both licensed and unlicensed bands. The tight integration of both technologies is seen as crucial for supporting the unrelenting growth in data traffic. In view of this, we developed a  cross-system learning framework aiming at optimizing the long-term performance of  SCBSs, in which delay-tolerant traffic is steered towards WiFi. Our approach is totally distributed with low signalling overhead, and shows significant improvements in terms of cell-edge UE throughput, especially in high load conditions. In our future investigations, we will extend the current formulation to the case of backhaul sharing, which is also gaining significant importance.

\end{document}